\documentclass[12pt]{article}

\begin{document}

\setcounter{page}{0}

\begin{titlepage}

\title{Quantum Spherical Spin Glass.  Supersymmetry and Annealing}

\author{Pedro Castro Menezes \\
           Alba Theumann\\
     Instituto de F\'{\i}sica, Universidade Federal do Rio Grande do Sul  \\
        Av. Bento Gon\c{c}alves 9500, 91501-970 Porto Alegre, RS, Brasil}

\date{}

\maketitle

\thispagestyle{empty}

\newpage

\begin{abstract}
\normalsize We show that the effective action of the quantum
spherical spin glass is invariant under a generalized form of the
Becchi-Rouet-Stora-Tyutin (BRST) supersymmetry. The Ward identities
associated to this invariance indicate that the spin glass order
parameter must vanish, and as a result the annealed average is exact
in this model. We present new results for the free energy, entropy
and specific heat. Due to quantum effects the entropy remains finite
and the specific heat vanishes at zero temperature. Results for the
phase diagram coincide with those obtained by different formalisms.
At zero temperature we derive the scaling behaviour with frequency
of the dynamical susceptibility .

\end{abstract}
\vspace*{0.5cm}

\noindent PACS numbers: 75.10.Nr; 75.10.Jm ; 64.70.Pf

\end{titlepage}


\setcounter{page}{1}

\section*{1.\ Introduction}
In a previous publication\cite{alba} the classical spherical spin
glass was analyzed by using supersymmetry
methods\cite{efetov,ramond} and it was pointed out that the model
was "anomalous" because the order parameter $ q$ that measures the
replica's overlap was vanishing and the configurational average of
any number of replicas gave the correct known
result\cite{kosterlitz}, then the annealed approximation with only
one replica was also correct. More recent work on the calculation
of the complexity, or logarithm of the density of states of a
certain energy, of the p-spherical spin glass, $p \ge 3 $,
\cite{annibale,cavagna} and of the Sherrington-Kirkpatrick model
\cite{crisanti} shows that the annealed average gives the correct
answer when the action is invariant under the generalized
Becchi-Rouet-Stora-Tyutin (BRST)
supersymmetry\cite{BRST} . \\
In the present paper we first demonstrate how the generalized BRST
supersymmetry transformation applies to the functional integral
formulation of the quantum spherical spin-glass.We also discuss how
this internal invariance leads to Ward identities that result in the
vanishing of the replica overlap $q$.This result had been obtained
previously for the classical spin-glass\cite{alba} and justifies the
performance of the annealed configurational
average\cite{annibale,cavagna,crisanti} also in the quantum
spherical spin-glass. The model was proposed earlier in
ref.\cite{shukla} and was solved by using the distribution of
eigenvalues of a gaussian random matrix. Later, the interest in this
model was renewed because it corresponds to the infinite-$M$ limit
of the spin-glass of $M$-components quantum rotors and it was
studied\cite{ye} by using the replica method in a replica symmetric
theory. Both theories give the same phase diagram, although
thermodynamic quantities as free energy and specific heat are not
discussed. On the other hand, the spin-glass model of quantum rotors
is equivalent to the Ising spin-glass model in a transverse
field\cite{sachdev}. A formulation for quantum spherical spins with
a different temporal dependence was presented for multispin
interactions\cite{theo}.Critical properties were analyzed in
ref.\cite{vojta} and applications to physical systems were presented
in ref.\cite{tu,mauricio}. All these papers coincide in that the
quantum spherical model presents a finite entropy and vanishing
specific heat at low temperature, while the classical model presents
a negative infinite entropy and constant specific heat.\\

  We study the quantum spherical spin-glass in the annealed
average by using a functional integral formulation derived by using
Feynman's prescription\cite{negele,feynman}. In sect.2 we present
the general formalism and discuss the supersymmetry transformation
that leaves invariant the action together with the corresponding
Ward identities\cite{amit}, while in sect.3 we discuss the
results.We are careful to normalize the free energy so as to recover
a finite classical limit when the rotors moment of inertia $ I
\rightarrow \infty $. In this way the phase diagram, that is
calculated from the spherical condition, coincide with previous
studies\cite{shukla,ye}.In the classical limit $ I \rightarrow
\infty $ the entropy diverges at $T=0$ in the typical fashion of the
classical spherical model\cite{alba,kosterlitz}, with the important
difference that it remains finite in the quantum regime for finite
values of the rotors moment of inertia. The specific heat shows the
usual discontinuity at the critical temperature but it vanishes at
$T=0$ in the quantum regime. We derive scaling laws for the
dynamical susceptibility, in agreement with previous
results\cite{ye}. We present in the Appendix some useful relations
among Grassmann variables without proof, while we refer the reader
to the original work\cite{efetov} for rigorous results.\\
\section*{2.\ Supersymmetry and Annealing}
We consider a spin glass of quantum rotors\cite{ye} with moment of
inertia $I$ in the spherical limit with Hamiltonian
\begin{eqnarray}
{\cal H}_{SG}+\mu \displaystyle \sum_{i}S_{i}^2 = \frac{1}{2I}
\displaystyle \sum_{i} P_{i}^2 - \displaystyle \frac{1}{2}
\sum_{i,j}J_{ij}S_{i}S_{j} + \mu \displaystyle \sum_{i}S_{i}^2
\label{1}
\end{eqnarray}
\noindent
where the spin variables at each site are continuous $ -\infty < S_{i} < \infty $
and we introduced
the canonical momentum $ P_{i}$ with commutation rules:\\
\begin{eqnarray}
[S_{j},P_{k}] = i \delta_{j,k} \label{2}
\end{eqnarray}
\noindent The sum  in eq.(\ref{1})runs over sites $ i,j =1..N $. The
coupling $J_{ij}$ in eq.(\ref{1}) is an independent random variable
with the gaussian distribution
\begin{eqnarray}
P(J_{ij})=e^{-J_{ij}^{2}\frac{N}{J^{2}}}\sqrt{\frac{N}{\pi J^{2}}}
\label{3}
\end{eqnarray}
while the chemical potential $\mu$ is a Lagrange multiplier that
insures the mean spherical condition
\begin{eqnarray}
-\frac{\partial\langle ln{\cal Z}\rangle}{\partial(\mu)}=
\displaystyle \sum_{i}\int_0^{\beta} d\tau \langle S_{i}^2 \rangle =
\beta N \label{4}
\end{eqnarray}
and $ \beta = 1/T $ is the inverse temperature. We work in units
where the Boltzmann constant $ k_B = \hbar= 1$ and ${\cal Q}$ is the
quantum partition function
\begin{eqnarray}
&&{\cal Z}= Tre^{-\beta({\cal H}_{SG}+\mu \sum_{i}S_{i}^2 )}
\label{5}
\end{eqnarray}
 The partition function at zero field can be expressed as a functional
integral\cite{negele,feynman}\\
\begin{eqnarray}
{\cal Z}= &&\displaystyle \int \prod_{i}{\cal D} S_{i}
\exp{\left(-\cal{ A_{O} - A_{SG}} \right)} \label{6}
\end{eqnarray}
where the non interacting action $ \cal A_{O} $ is given by
\begin{eqnarray}
{ \cal A_{O}} =\int_{0}^{\beta} d \tau  \displaystyle
\sum_{i}\left(\frac{I}{2} \left(\frac{\partial S_{i}}{\partial
\tau}\right)^{2} + \mu S_{i}^2(\tau)-H_{i}(\tau)S_{i}(\tau)\right)
\label{7}
\end{eqnarray}
and the interacting part
\begin{eqnarray}
{ \cal A_{SG}} =\frac{1}{2}
\displaystyle\sum_{i,j}J_{ij}\int_{0}^{\beta} d \tau
S_{i}(\tau)S_{f}(\tau) \label{8}
\end{eqnarray}
After performing a Fourier transformation in time and taking into
account the reality of the fields $ S_{i}(\tau)$ the generating
functional can be written in terms of the real components $ R_{i}(n)
$ of the Fourier transforms $
S_{i}(\omega_{n})=S_{i}^{\ast}(-\omega_{n})$, where $\omega_{n}=
\frac{2\pi n}{\beta }$ are boson Matsubara's frequencies. As it was
shown in \cite{alba} the correlation functions can be obtained from
the generating functional for two replicas
\begin{eqnarray}
{\cal W}(H)=\displaystyle \prod_{n\ge 0}\Vert \Gamma_{ij}(n)\Vert
\int\prod_{i\alpha n\ge 0} dR_{i\alpha}(n)\nonumber
\\
\exp{\left(-\sum_{\alpha n\ge
0}\sum_{ij}\Gamma_{ij}(n)R_{i\alpha}(n)R_{j\alpha}(n)
+\sum_{i\alpha}
H_{i\alpha}(n)R_{i\alpha}(n)\right)}\nonumber\\
\label{9}
\end{eqnarray}
where the index $\alpha = 1,2$, the $ H_{i\alpha}(n) $ are auxiliary
fields and
\begin{eqnarray}
\Gamma_{ij}(n)=(I\beta\omega_n^2 + 2\beta\mu)\delta_{ij} - J_{ij}
\label{10}
\end{eqnarray}
By using the results in the Appendix , the determinant $ \Vert
\Gamma_{ij}(n)\Vert $ may be expressed with the help of auxiliary
Grassmann fields $ \chi^\ast, \chi$ and $ \cal W $ can be written
\begin{eqnarray}
{\cal W}(H_\alpha ,\gamma)=\displaystyle \prod_{n\ge 0}\int\prod_{i
n\ge 0} d\chi_i^\ast(n)d\chi_i(n)\prod_{\alpha} dR_{i\alpha}(n)
\nonumber\\
\exp-\biggl[\displaystyle\sum_{n\ge
0}\Bigl[\sum_{ij}\Gamma_{ij}(n)\Bigl(\frac{\chi^{*}_{i}(n)\chi_{j}(n)+
\chi^{*}_{j}(n)\chi_{i}(n)} {2}+
\sum_\alpha R_{i\alpha}(n)R_{j\alpha}(n)\Bigr)\nonumber\\
+\sum_i\left(\frac{\gamma_i^\ast(n)\chi_i(n)+\chi_i^\ast(n)\gamma_i(n)}{2}+
\sum_\alpha H_{i\alpha}(n)R_{i\alpha}(n)\right)\Bigr]\biggr]\nonumber\\
\label{11}
\end{eqnarray}
where $\chi^{*}_{i}(n),\chi_{i}(n)$ are complex anticommuting
Grassmann variables while the $ R_{i,\alpha}(n), \alpha=1,2 $ are
real commuting variable and we introduced two extra Grassmann
auxiliary fields $\gamma_i $ and $ \gamma^\ast_i $. As the
$\Gamma_{ij}$ are symmetric, when the auxiliary fields are set equal
to zero the functional is invariant under the supersymmetry
transformation
\begin{eqnarray}
\chi_{i}^{\prime\ast}(n)/\sqrt2=\chi_{i}^\ast(n)/\sqrt2+
\epsilon^\ast\sum_{\alpha}R_{i,\alpha}(n)/\sqrt2\nonumber\\
\chi_{i}^{\prime}(n)/\sqrt2=\chi_{i}(n)/\sqrt2+
\epsilon\sum_{\alpha}R_{i,\alpha}(n)/\sqrt2\nonumber\\
R_{i,\alpha}(n)^{\prime}=R_{i,\alpha}(n)-
\frac{\epsilon^\ast\chi_{i}(n)}{2}
+\frac{\epsilon\chi_{i}^{\ast}(n)}{2} \label{12}
\end{eqnarray}
where $\epsilon$ is a complex Grassmann variable and we adopt the
convention for complex conjugation\cite{efetov}
$\epsilon^{\ast\ast}=-\epsilon$,
$(\epsilon\chi_{i}^\ast)^{\ast}=\epsilon^{\ast}\chi_{i}^{\ast\ast}=
-\epsilon^\ast\chi_{i}$, then $R_{i,\alpha}(n)^{\prime}$ is
effectively a real variable. The transformation differs
infinitesimally from unity and $ \epsilon^\ast\epsilon $  can be
neglected, although this concept is somehow meaningless in the case
of Grassmann variables. Now it is convenient to switch to the
superalgebra notation\cite{efetov} of the Appendix and to introduce
the supervectors for the auxiliary fields
\begin{eqnarray}
\underline {J}_i=\pmatrix{H_{i1}\cr
H_{i2}\cr\gamma_i/\sqrt2\cr\gamma_i^\ast/\sqrt2\cr}\nonumber\\
\underline{SJ}_i^\dagger=
\pmatrix{H_{i1}&H_{i2}&\gamma_i^\ast/\sqrt2&-\gamma_i/\sqrt2}
\label{13}
\end{eqnarray}
and for the field variables
\begin{eqnarray}
\underline {\varphi}_i=\pmatrix{R_{i1}\cr
R_{i2}\cr\chi_i/\sqrt2\cr\chi_i^\ast/\sqrt2\cr}\nonumber\\
\underline{S\varphi}_i^\dagger=
\pmatrix{R_{i1}&R_{i2}&\chi_i^\ast/\sqrt2&-\chi_i/\sqrt2} \label{14}
\end{eqnarray}
In this way the functional $\cal W$ in equation(\ref{11}) may be
written in compact form
\begin{eqnarray}
{\cal W}(J)=\displaystyle \prod_{n\ge 0}\int\prod_{i n\ge 0}
d\chi_i^\ast(n)d\chi_i(n)\prod_{\alpha} dR_{i\alpha}(n)
\nonumber\\
\exp\biggl[-1/2\displaystyle
\sum_{ij}\underline{S\varphi_i}^\dagger{\bf
M}_{ij}\underline\varphi_i + \displaystyle
\sum_i\underline{SJ}_i^\dagger\underline {\varphi}_i\biggr]
\label{15}
\end{eqnarray}
that may be formally integrated to give
\begin{eqnarray}
{\cal W}(J)=\exp\biggl[1/2\displaystyle
\sum_{ijn}\underline{SJ_i}^{\dagger}{\bf Q}_{ij}\underline{J}_j
\biggr] \label{16}
\end{eqnarray}
with the 4x4 supermatrices
\begin{eqnarray}
\bf{Q}_{ij}=\pmatrix{q_{11}&q_{12}&\theta_{13}&\theta_{14}\cr
q_{21}&q_{22}&\theta_{23}&\theta_{24}\cr
\theta_{31}&\theta_{32}&v_{33}&0\cr
\theta_{41}&\theta_{42}&0&v_{44}\cr}_{ij}\label{17}
\end{eqnarray}
By symmetry we should have
\begin{eqnarray}
q_{11}=q_{22}=<R_{i\alpha}R_{i\alpha}>=v_{33}=v_{44}=<\chi^\ast
\chi> \nonumber\\
v_{34}=v_{43}=<\chi \chi> = <\chi^\ast\chi^\ast>=0 \label{18}
\end{eqnarray}
To derive Ward identities\cite{amit} we consider that, as the system
is invariant when the fields are subject to the supersymmetry
transformation in equation(\ref{12})
\begin{eqnarray}
\underline{\varphi}_i^\prime={\bf A}\underline{\varphi}_i \nonumber
\end{eqnarray}
where
\begin{eqnarray}
 {\bf A}=\pmatrix{1&0&-\epsilon^\ast/\sqrt2&\epsilon/\sqrt2\cr
0&1& -\epsilon^\ast/\sqrt2&\epsilon/\sqrt2\cr \epsilon/\sqrt2&
\epsilon/\sqrt2&1&0\cr\epsilon^\ast/\sqrt2&
\epsilon^\ast/\sqrt2&0&1\cr}= \bf1 + \bf{D}\label{19}
\end{eqnarray}
then it should be also invariant when we transform the external
fields \cite{amit}  in equation(\ref{16})
\begin{eqnarray}
\underline{J}^\prime={\bf A}\underline{J} \nonumber\\
\underline{SJ}^{\dagger\prime}=\underline{SJ}^\dagger{\bf
SA}^\dagger\label{20}
\end{eqnarray}
Where the super-adjoint ${\bf SA}^\dagger $ is defined in the
Appendix. This condition leads to
\begin{eqnarray}
\bf{SD}^\dagger\bf{Q}+\bf{Q}\bf{D}=0\label{21}
\end{eqnarray}
from where we deduce
\begin{eqnarray}
v_{33}=q_{11}+q_{12}\nonumber\\
v_{44}=q_{22}+q_{21}\label{22}
\end{eqnarray}
and from equation(\ref{18}) we obtain for the overlap between two
replicas
\begin{equation}
q_{12}=q_{21}=0\label{23}
\end{equation}
The spin-glass order parameter $ q = q_{12}$ and it vanishes
identically from equation(\ref{19}), thus making the annealed
average exact in the quantum spherical spin-glass as it was
discussed in\cite{alba} for the classical spherical spin-glass.
\section*{3.\ Results}
When performing the annealed average we take the configurational
average of $\cal Z $ in equation(\ref{6}) over the random
variables $J_{ij}$ and after splitting the quadratic term with a
gaussian integration we obtain the result
\begin{eqnarray}
\left\langle {\cal Z}\right\rangle _{ca}= \int {\cal D}Q(\tau
-\tau^{\prime})\exp{\left[-N\left(\frac{J^{2}}{4}\int_{0}^{\beta}\int_{0}^{\beta}
 d \tau d \tau^{\prime}Q(\tau -\tau^{\prime})^{2} -
 \Lambda \right)\right]}
 \label{24}
\end{eqnarray}
where we assumed time translational invariance and
\begin{eqnarray}
e^{\Lambda}=\int{\cal D}S(\tau)\exp{\left[-{\cal A_{O}}
+\frac{J^{2}}{2}\int_{0}^{\beta}\int_{0}^{\beta}d \tau d
\tau^{\prime} Q(\tau- \tau^{\prime})S(\tau)S(\tau^{\prime})\right]}
\label{25}
\end{eqnarray}
A steepest descent calculation in eq.(\ref{24}) gives
\begin{eqnarray}
Q(\tau-\tau^{\prime})= \frac{1}{N}\sum_{i} <
S_{i}(\tau)S_{i}(\tau^{\prime})> \label{26}
\end{eqnarray}
The partition function in eq.(\ref{24}) is solved by a time Fourier
transformation with the result
\begin{eqnarray}
\left\langle {\cal Z}\right\rangle _{ca}={\mathcal
N}^{N}\prod_{n=0}^{\infty}\int dQ(\omega_{n})\exp\{-N[\frac{(\beta
J)^{2}}{2} Q(\omega_{n})^{2}-\frac{\beta
H^{*}(\omega_{n})H(\omega_{n})}
{\beta I{\omega_{n}}^{2}+2\beta\mu-2(\beta J)^2 R(\omega_{n})}  \nonumber \\
 + \ln\left(\beta I \omega_{n}^{2}+2\beta \mu-(\beta J )^{2}
  Q(\omega_{n})\right)]\} \nonumber \\
\label{27}
\end{eqnarray}
where $\omega_{n} = \frac{2\pi n}{\beta}$ is the boson Matsubara
frequency and $ Q(\omega_n) $ is the real part of the Fourier
transform of $ Q(\tau) $. A steepest descent evaluation of the
integral in eq.(\ref{27}) gives, for $H(\omega_{n})=0 $
\begin{eqnarray}
Q(\omega_{n})=[\beta I \omega_{n}^{2}+2\beta \mu-(\beta
J)^2Q(\omega_{n})]^{-1}
 \label{28}
\end{eqnarray}
with solution
\begin{eqnarray}
Q(\omega_{n})=\frac{\beta I \omega_{n}^{2}+2\beta \mu- \sqrt{(\beta
I \omega_{n}^{2}+2\beta \mu)^2-4(\beta J )^2}}{2(\beta J)^2}
\label{29}
\end{eqnarray}
while the mean spherical condition in eq.(\ref{4}) becomes
\begin{eqnarray}
\sum_{-\infty}^{\infty}Q(\omega_{n})=1 \label{30}
\end{eqnarray}
Introducing eq.(\ref{29}) into eq.(\ref{27})we obtain for the free
energy per site at the saddle point
\begin{eqnarray}
\beta F= \displaystyle
\sum_{-\infty}^{\infty}\left(\frac{1}{4}(\beta J)^2 Q(\omega_{n})^2
-\frac{1}{2} \ln\left(Q(\omega_{n})\right)\right)-\beta \mu
-\ln\left(\mathcal N \right) \label{31}
\end{eqnarray}
The normalization constant $ {\mathcal N} = \frac{1}{\sqrt{IJ}} $ is
determined so that we recover a finite limit for the free energy
when $ I \rightarrow \infty $ . Following standard
procedures\cite{negele} we convert sums over frequencies into
integrals with the result for the mean spherical condition
\begin{eqnarray}
\displaystyle
\int_{L_{-}}^{L_{+}}dy\sqrt{(L_{+}^2-y^2)(y^2-L_{-}^2)}\coth(\frac{\beta
y}{2\sqrt{I}})= 2\pi J^{2}\sqrt{I} \label{32}
\end{eqnarray}
where
\begin{eqnarray}
L_{\pm}^2 = 2\mu \pm 2J \label{33}
\end{eqnarray}
and we obtain for the free energy
\begin{eqnarray}
\beta F=\frac{1}{\pi J^2}\displaystyle \int_{L_{-}}^{L_{+}}dy
y\ln\sinh\left(\frac{\beta
y}{2\sqrt{I}}\right)\sqrt{4J^2-(2\mu-y^2)^2}- \beta \mu
-\ln{ \mathcal N} \nonumber \\
\label{34}
\end{eqnarray}
 In the classical limit $ I\rightarrow\infty $ the integrals in
eq.(\ref{32}) and eq.(\ref{34}) can be performed exactly in terms of
hypergeometric functions with the result
\begin{eqnarray}
\frac{2\mu_{class}}{J}= \beta J + \frac{1}{\beta J}\label{35}
\end{eqnarray}
\begin{eqnarray}
\beta F_{class}=\frac{1}{2} \ln\left[\beta\mu +
\sqrt{(\beta\mu)^2-(\beta J)^2}\right] + \frac{1}{4}
\frac{\beta\mu - \sqrt{(\beta\mu)^2-(\beta J)^2}}{\beta\mu +
\sqrt{(\beta\mu)^2-(\beta J)^2}}-\mu\beta- \nonumber
\\ +\ln{\frac{\sqrt{\beta J}}{2}} \nonumber \\
 \label{36}
\end{eqnarray}
The expression in eq.(\ref{35}) coincide with the results for the
classical spherical spin glass in ref.(\cite{alba,kosterlitz}), but
the free energy in eq.(\ref{36}) differs due to the last logarithmic
contribution. This originates in the choice of the normalization
constant $\mathcal N $. If we had chosen $\mathcal N $ following
Feynman's prescription\cite{negele,feynman} we would have obtained
unphysical results in the quantum regime for finite values of $I$,
like a negative specific heat at low temperatures \cite{pedro}. The
chemical potential $\mu $ decreases with temperature until it
reaches the value $ \mu_c = J $\cite{kac} at the critical
temperature $ T_{c} $, for lower temperatures the spherical
condition no longer holds and $ \mu $ sticks to this value. We show
in fig.1 the solution for $ \mu $ as a function of $ T $ for
different values of $ I $. The critical temperature is obtained by
setting $\mu = \mu_c =J $ in eq.(\ref{35}). We show in fig.2 the
phase diagram for $ T_{c} $ as a  function of $ I $. $T_c $
decreases with $ 1/I $ and vanishes at the critical value $I_c =
\frac{16}{9 \pi^2 J} $. This result coincides with previous
references\cite{shukla,ye}. Close to this value we find $T_c^2 =
\sqrt{\frac{J(I-I_c)}{3I_c^2}}$. We show in fig.3, fig.4, fig.5
results for the free energy, entropy and specific heat as functions
of temperature for different values of $ I $. We observe that the
entropy remains finite when $ T \rightarrow 0 $   for any finite
value of $I$, while we recover  the usual divergence $ S \rightarrow
-\infty$ in the classical spherical spin glass
model\cite{alba,kosterlitz} when $ I \rightarrow \infty $. The
specific heat vanishes at $ T=0 $ for finite values of $I$, as
expected in quantum systems\cite{theo,vojta,tu,mauricio}. The
dynamic susceptibility is the response to the field $ H(\omega_{n})$
and we obtain from eq.(\ref{27})
\begin{eqnarray}
\chi(\omega_n)=\frac{\omega_n^{2}I +2 \mu-\sqrt{(\omega_n^{2}I +2
\mu)^2-4J^2}}{2J^2} \label{37}
\end{eqnarray}
The static susceptibility
\begin{eqnarray}
\chi(0)=\frac{ \mu-\sqrt{ \mu^2-J^2}}{J^2} \label{38}
\end{eqnarray}
satisfies Curie law for high temperatures and exhibits the usual
cusp at the critical temperature. It is shown in fig.6 for different
values of $ I $. The dynamic susceptibility at $ T=0 $ is obtained
by setting $i\omega_n \rightarrow \omega+i\delta $ in
eq.(\ref{37}).For $\Delta= \mu-J \geq 0$ and $ I = I_c $ we obtain
that the $ Im\chi(\omega) $ satisfies the scaling relation
\begin{eqnarray}
Im\chi(\omega) =sgn{\omega}|\omega|^{\mu^\prime} (\sqrt{I_c}/J)
\Phi(\omega\sqrt{\frac{I_c}{J^2 \Delta}})\label{39}
\end{eqnarray}
with the exponent $\mu^\prime =1$ and the scaling function $
\Phi(x)=\sqrt{1-x^{-2}}$, what agrees with the results of
ref.(\cite{ye}).
\section*{4.\ Conclusions}
The quantum spherical spin glass was studied within the annealed
average, that is exact  due to the internal BRST supersymmetry of
the model,as it was discussed in sect.1. Our results for the phase
diagram coincide with those obtained by Ye et al\cite{ye} with the
replica method and by Shukla et al \cite{shukla} using the
semicircular law. However, a substancial difference exists between
our results for the free energy, entropy and specific heat and those
presented by Shukla et al\cite{shukla}. The results obtained with
quantum functional integrals methods are very sensitive to the use
of the correct normalization and measure \cite{negele,feynman} and
we set the normalization constant in our expression for the free
energy such as to recover a finite limit when $ I \rightarrow\infty
$. As a result we obtain a finite entropy at low temperatures and a
vanishing specific heat for finite values of the rotors moment of
inertia $I$, as it is expected in quantum spherical
models\cite{theo,vojta,tu,mauricio}. In the limit $ I\rightarrow
\infty$  we recover the known negative,infinite entropy and a
constant specific heat at $T=0 $. Also at $T=0$ we derive the
scaling behaviour for the dynamical susceptibility that agrees with
Ye et al\cite{ye}.
\newpage
\section*{5.\ Appendix}
We refer here to references \cite{efetov,ramond} to give a concise
and brief description of Grassmann variables and superalgebra. We
indicate  complex Grassmann variables by Greek letters, and these
anticommuting variables satisfy
\begin{eqnarray}
\left[\chi, \gamma \right]_+=0 \hspace{1cm}   \chi^2={\chi^{\ast}}^2=0 \nonumber\\
{\chi^\ast}^\ast = -\chi \nonumber\\
(\chi \gamma)^\ast =\chi^\ast \gamma^\ast \nonumber\\
(\chi^\ast \chi)^\ast ={\chi^\ast}^\ast\chi^\ast = -\chi \chi^\ast=
\chi^\ast \chi\label{40}
\end{eqnarray}
and the integrals
\begin{eqnarray}
\int d\chi^\ast \chi^\ast  =\int d\chi \chi=1 \nonumber \\
\int d\chi =\int d\chi^\ast=0
\end{eqnarray}
what leads to
\begin{eqnarray}
\int \prod_i d\chi^\ast_i d\chi_i \exp{\sum_{ij}\chi_i^\ast
M_{ij}\chi_j} = \Vert M_{ij}\Vert \label{41}
\end{eqnarray}
A four-component supervector ${\underline \varphi}$ and a $4x4$
super matrix {\bf M }have commuting and Grassmann components
\begin{eqnarray}
{\underline \varphi} =\pmatrix{R_1\cr R_2\cr \chi/\sqrt2\cr
\chi^\ast/\sqrt2}\nonumber\\
{\bf M}= \pmatrix{\underline a & \underline\sigma \cr \underline\rho
& \underline b} \label{42}
\end{eqnarray}
where $\underline a $ and $ \underline b $ are $ 2x2$ matrices of
commuting elements while $\underline\sigma $ and $\underline\rho $
are $2x2$ matrices of Grassmann elements. Taking into account the
change of sign involved in complex conjugation from eq(\ref{40}) the
super adjoint must be defined
\begin{eqnarray}
S{\underline \varphi}^\dagger =\pmatrix{R_1 & R_2 & \chi^\ast/\sqrt2
&
-\chi/\sqrt2}\nonumber\\
S{\bf M}^\dagger= \pmatrix{\underline a^\dagger &
\underline\rho^\dagger \cr -\underline\sigma^\dagger & \underline
b^\dagger}  \label{43}
\end{eqnarray}
\newpage
\section*{6.\ Figure Captions}
Fig.1 Chemical potential $\mu$ as a function of temperature for
different values of the moment of inertia $ I =0.19 J $(dash-dot),
$I=J $
(dash), $ I=\infty $(continuous) .\\

Fig.2 Phase diagram and critical line $ T_c(1/I) $ separating the
paramagnetic from the spin glass phase in the $T vs 1/I $ plane.\\

Fig.3 Free energy as a function of temperature for different values
of the moment of inertia $ I =0.4 J $(dash), $I=J$
(dash-dot-dot),$I=5J $(dash-dot),$ I=10J $(dash-dot-dot)
 $I=\infty$ (continuous) .\\

 Fig.4 Entropy as a function of temperature for different
values of the moment of inertia $ I = J (dash), I= 5J
(dash-dot-dot),
I=10J(dash-dot),I=\infty(continuous) $.\\

Fig.5 Specific heat as a function of temperature for different
values of the moment of inertia $ I =0.4 J$ (dash),$ I= J$
(dash-dot),
$I=10J$(dash-dot-dot),$I=\infty$(continuous) .\\

Fig.6 Static susceptibility as a function of temperature for $
I=0.4J$(dash), $I=J$ (dash-dot),$I=\infty $(continuous) .\\

\newpage

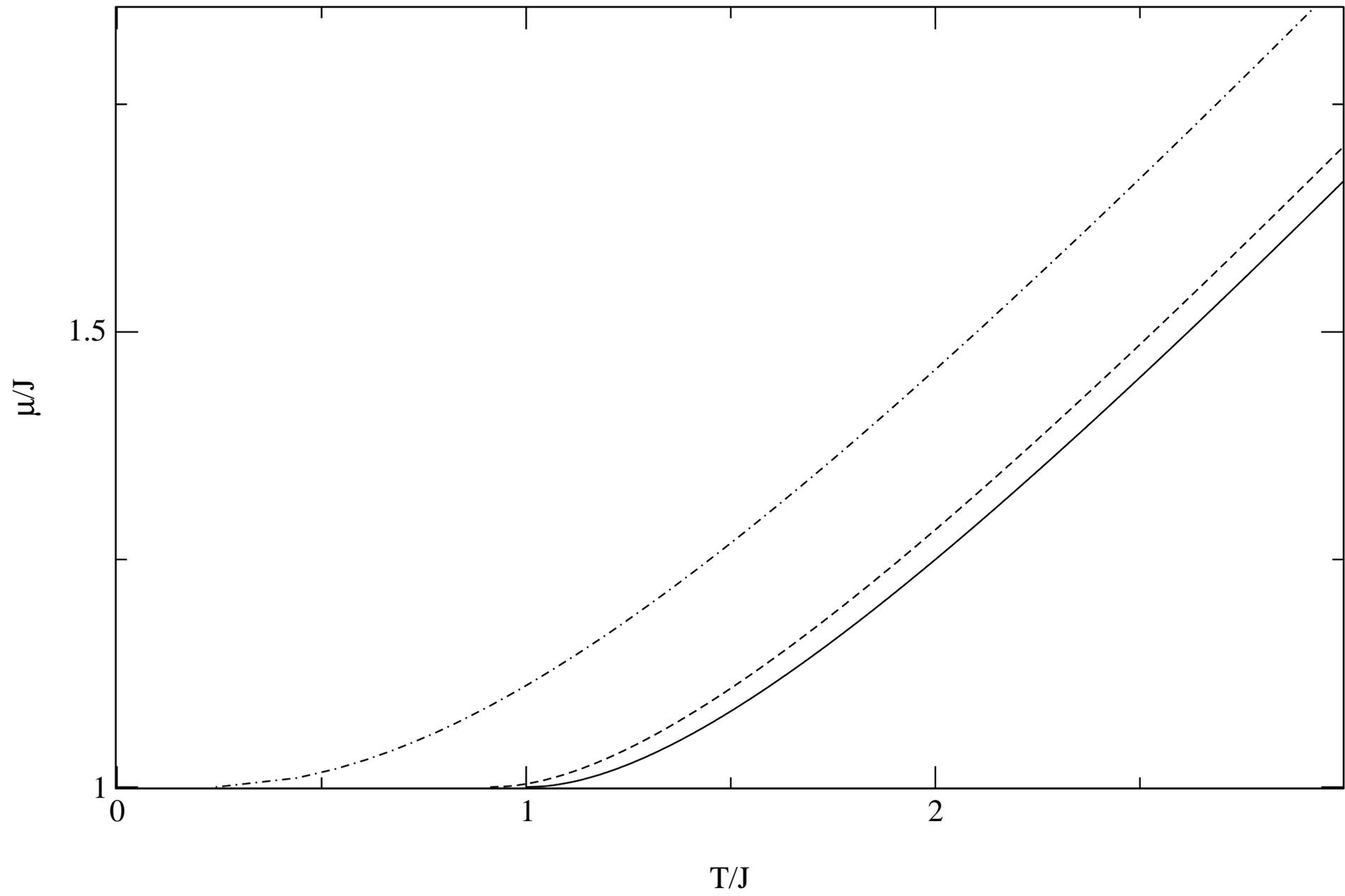
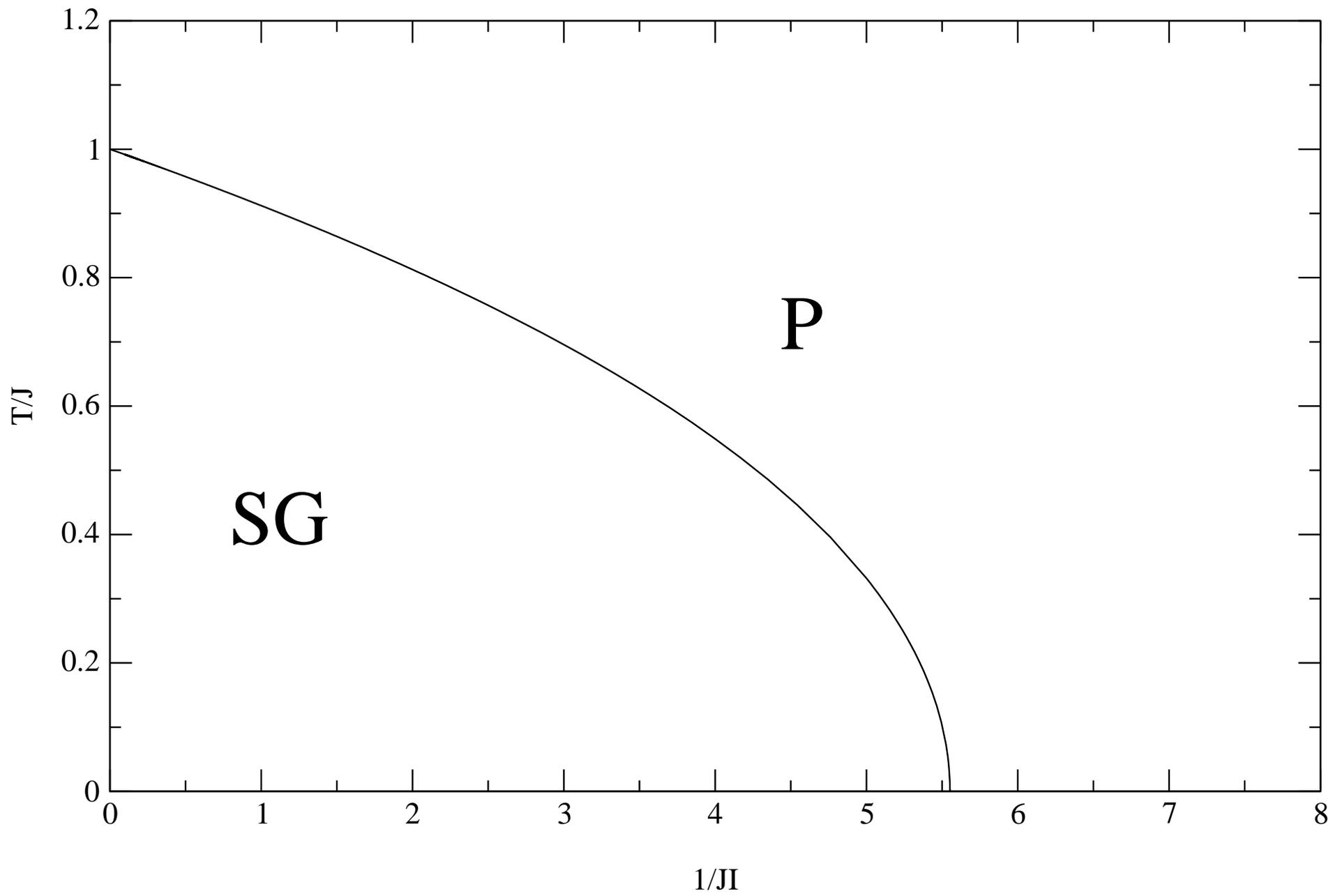
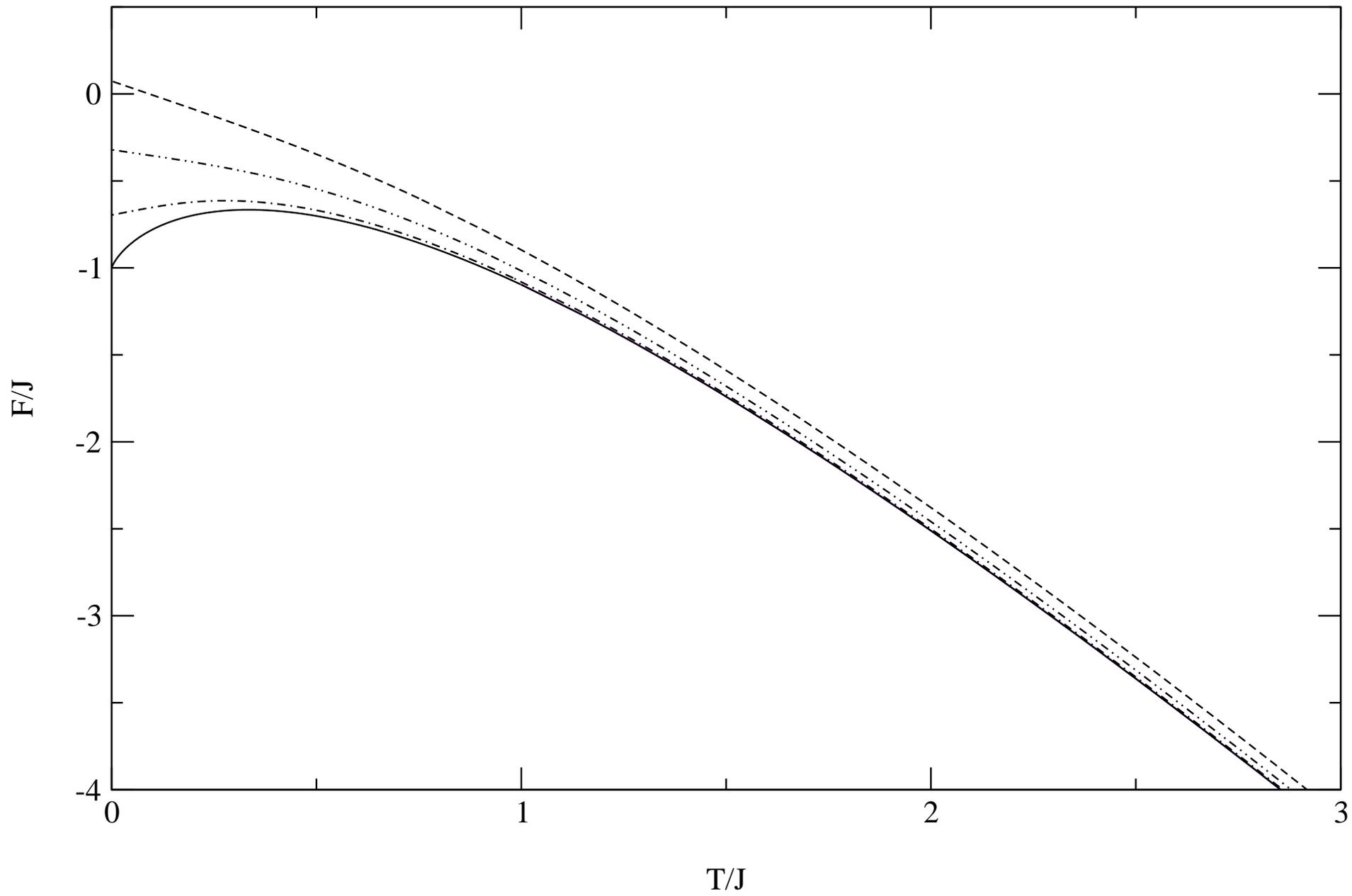
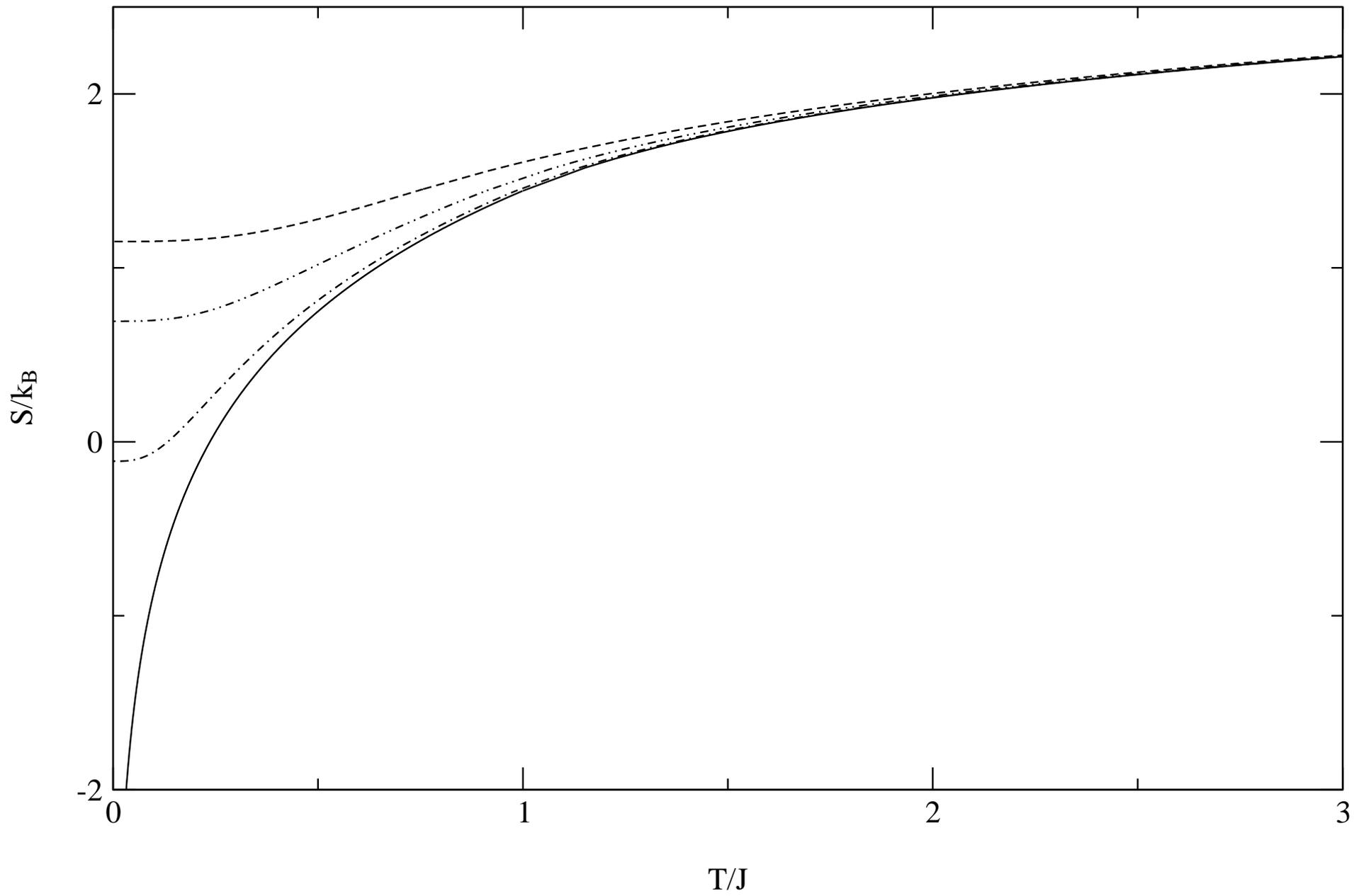
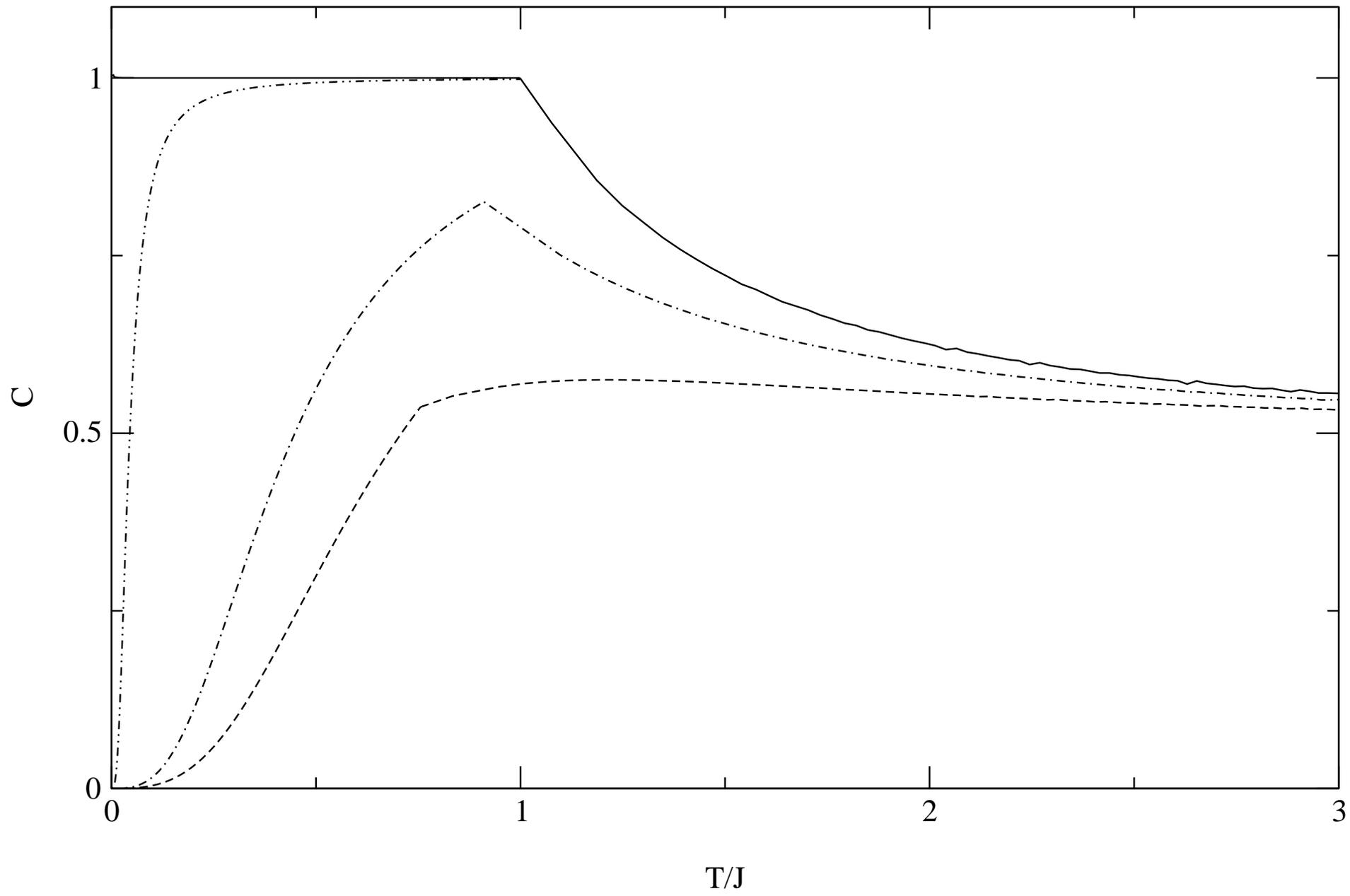
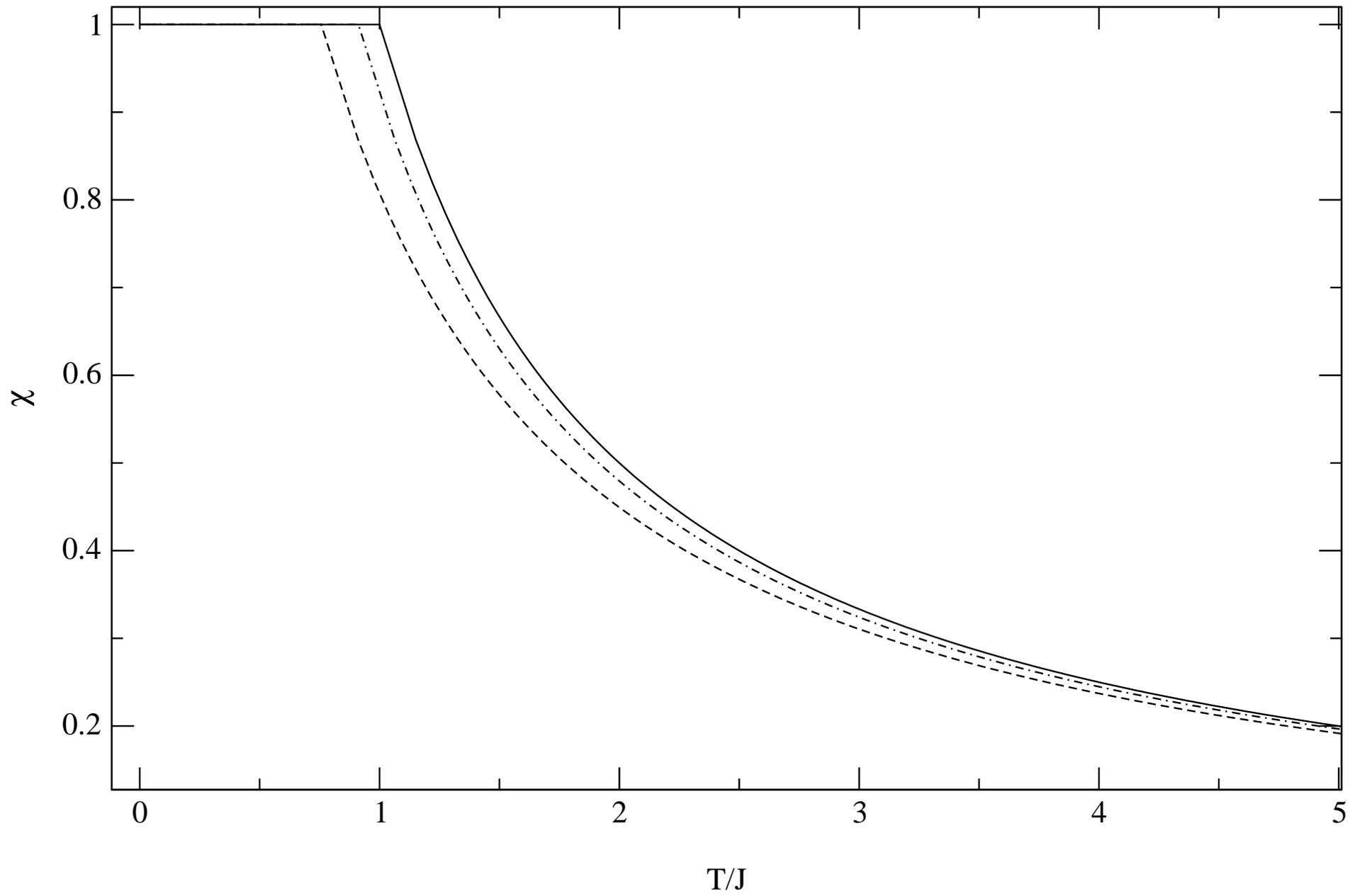
\end{document}